\documentclass[12pt]{article}
\usepackage{amsfonts}

\usepackage{graphicx}
\usepackage{amsmath}

\input{tcilatex}

\begin{document}

\baselineskip16pt

\begin{titlepage}

\  \\

\vspace{16 mm}

\begin{center}
{\Large \bf Tachyon Condensation in Large Magnetic Fields}\\
\vspace{2mm}
{\Large \bf with}\\
\vspace{2mm}
{\Large \bf Background Independent String Field Theory}\\

\vspace{3mm}

\end{center}

\vspace{8 mm}

\begin{center}

Lorenzo Cornalba

\vspace{3mm}

{\small \' Ecole Normale Sup\'erieure} \\
{\small Paris, France} \\
{\small  cornalba@lpt.ens.fr} \\

\vspace{5mm}

October 2000

\vspace{5mm}

\end{center}

\begin{abstract}

{ \it

We discuss the problem of tachyon condensation in the framework
of background independent open
string field theory. We show, in particular, that the computation of
the string field theory action simplifies considerably if one looks at closed string
backgrounds with a large $B_{ab}$ field, and can be carried out
exactly for a generic tachyon profile. We confirm previous results on
the form of the exact tachyon potential, and we find, within this framework,
solitonic solutions which correspond to lower dimensional unstable branes.

}
\end{abstract}

\end{titlepage}
\newpage

\section{Introduction}

The problem of tachyon condensation on unstable branes has received lately a
lot of attention \cite
{KMM,SZ1,SZ2,MT,HK,KJMT,MSZ,GSS,GS,S1,S2,S3,S4,GMS1,GMS2,HKLM}, and
considerable progress has been made following the initial conjectures of Sen
(see \cite{S3,S4}and references therein). The problem is most easily studied
in the context of bosonic open string physics, where the relative simplicity
of the theory has led both to numerical as well as exacts checks of the
conjecture.

Questions on tachyon condensation are most naturally addressed within string
field theory\cite{W,W1}, since we are interested in processes which connect
two distinct vacua of the theory, and which are by their very nature
off--shell non--perturbative phenomena. More precisely, Sen's conjecture
states that we should find a variety of stationary points of the string
field theory action, corresponding to branes of various dimensions.
Moreover, the value of the action at these stationary points should equal
the energy of the corresponding brane configurations. Finally the minimum of
the string field theory action should be given by a unique transitionally
invariant vacuum which contains no open strings at all.

>From a practical point of view, two are the questions that need to be
answered. First of all, one needs to compute the tachyon potential, and show
that the difference in energy between the maximum and the minimum
corresponds to the energy of a single $25$--brane. Second, on needs to find,
within the string field theory, classical solitons which correspond to
branes of lower dimensions.

These questions have been mostly addressed within the framework of Witten's
cubic Chern--Simons string field theory \cite{W}, and have been studied in
the two different closed string backgrounds, either with vanishing NSNS
two--form field $B_{ab}$, or with an infinitely large $B_{ab}$. In the first
case, most results are numerical \cite{KS,MT,SZ1,SZ2}, and have led, using
the level truncation scheme, to computations of the tachyon potential which
confirm Sen's conjecture to approximately $1\%$. Tensions of branes are more
difficult to obtain in this case, since the corresponding solitons are
expected to have a characteristic size of order $l_{s}$, and therefore all
derivative corrections to the string action will then contribute with the
same weight. Indeed, solitons can be found even restricting the action to
two derivatives \cite{HK,KJMT,MSZ}, but the energies obtained are usually
only qualitatively correct. Still in the $B_{ab}=0$ regime, an analytical
construction of the stable vacuum has been given in \cite{KP}. In the limit
of large $B_{ab}$ nothing new can be said about the tachyon potential, but
solitons corresponding to the various branes can be explicitly characterized
and energies can be computed exactly\cite{HKLM,GMS1}. This is the case since
the presence of a magnetic field introduces a new length scale which becomes
much larger then the effective string scale, and which controls the size of
the solitonic solutions. Therefore $\alpha ^{\prime }$ derivative
corrections to the action become irrelevant in the computation of the
energies. The unique information which is needed in order to compute
tensions exactly is the value of the tachyon potential at the extremum
points, which is assumed from the conjecture of Sen.

A different approach to the problems discussed above has been proposed in 
\cite{GSS,KMM}, and is based on an alternative background independent
formulation of open string field theory, which was originally proposed in 
\cite{FT} and then subsequently developed in \cite{W1,W2,SS1,SS2}. The basic
data is a two dimensional field theory defined on a disk string world--sheet 
$\Sigma $ (the disk being equipped, for definiteness, with the standard
metric). The action is conformal in the interior of $\Sigma $, and is the
one defining the closed string background. On the boundary, on the other
hand, one can generically insert any operator (of ghost number zero)\ built
from the fields of the theory and their derivatives. Every single boundary
operator will come with its own coupling constant $g$ and we should consider
the space of all coupling constants as the configuration space of string
field theory. The action functional $S\left( g\right) $ is then very simply
computed starting from the partition function $Z\left( g\right) $ of the
above theory, as well as from the $\beta $--function RG flow in the
configuration space of coupling constants $g$.

Already in the early work of \cite{FT} it was conjectured that a stable
vacuum could correspond to a breaking of Lorentz invariance from $SO\left(
1,25\right) $ to $SO\left( 2\right) \times SO\left( 1,23\right) $, or, in
modern language, to a decay from a $25$ to a $23$--brane. More recently, in 
\cite{GSS}, the authors compute, for generic tachyon profile and for
vanishing $B_{ab}$, the effective action up to two derivatives. In
particular, they are able to compute exactly the tachyon potential,\
precisely confirming Sen's conjecture. Moreover, in \cite{KMM} the authors
have shown that, if one works with the same closed string background with $%
B_{ab}=0$, and if one considers as a boundary operator a tachyon profile
which is quadratic in the space--time coordinates, then one can compute the
action to all orders in derivatives (based on \cite{W2}). Moreover one can
describe explicitly the solitons corresponding to the lower dimensional
branes, and compute exactly their energy.

In this letter we show that, if one considers the procedure of \cite{KMM,GSS}
but concentrates, as a closed string background, on the limit of large $%
B_{ab}$, then one can compute the action for generic tachyon profiles
extremely easily. We recover the expected form of the potential, and we
compute energies of solitonic solutions. All results agree with the known
expectations, and we make some contact with the usual formulation \cite
{HKLM,GMS2} of the problem in cubic open string field theory in the large $%
B_{ab}$ limit.

\section{Large Magnetic Field Limit}

We study a bosonic open string moving in Euclidean space--time, and we take
the metric to be, for simplicity, equal to $\delta _{ab}$. We will
concentrate, in particular, on the motion in two of the Euclidean
directions, with coordinates $x^{1}$ and $x^{2}$, and we shall collectively
call the other $24$ perpendicular coordinates $y_{\perp }$. We will finally
have, as a background field, a constant NSNS $2$--form potential in the $1$--%
$2$ direction given by 
\begin{equation*}
B_{12}=B.
\end{equation*}
Units will be such that $2\pi \alpha ^{\prime }=1$.

We consider, following \cite{W1,SS1,SS2}, string actions defined on the unit
disk $\Sigma \subset \mathbb{C}$ (endowed with the standard flat metric) of
the general form 
\begin{equation*}
I=I_{0}+\frac{1}{2\pi }\int_{\partial \Sigma }d\tau \,V.
\end{equation*}
In the above equation, the action $I_{0}$ is given by 
\begin{eqnarray*}
I_{0} &=&\frac{1}{2}\,\delta _{ab}\int_{\Sigma }dX^{a}\wedge \star
dX^{a}+iB\int_{\Sigma }dX^{1}\wedge dX^{2} \\
&&+\text{ terms in }Y_{\perp }\text{ and ghost fields,}
\end{eqnarray*}
and is the usual conformal action defined on the interior of the disk $%
\Sigma $ which describes the motion of the string in the chosen closed
string background. On the other hand, the operator $V$ inserted on the
boundary is a general linear combination 
\begin{equation*}
V=V_{i}g^{i},
\end{equation*}
where the operators $V_{i}$ span the set of all possible boundary operators
built only from the matter part of the theory\ (the most general
construction, which does not assume a decoupling of matter and ghosts, is
given in \cite{W1}, but we will not need it in the sequel).

In the rest of this letter, we will actually restrict our attention, as
indicated at the beginning of the section, to boundary operators $V_{i}$
built only from the matter fields $X^{a}$ and independent of the transverse $%
Y_{\perp }$. This means that we are considering physical configurations
which are invariant under translations in the transverse directions $%
y_{\perp }$.

The main result of \cite{SS2} is that, if one considers the partition
function

\begin{equation}
Z\left( g\right) =\int DX\;e^{-I(X)},  \label{eq10}
\end{equation}
then the classical background independent effective action which describes
open string boundary interactions is given by 
\begin{equation}
S=\left( 1+\beta ^{i}\frac{\partial }{\partial g^{i}}\right) Z,  \label{eq20}
\end{equation}
where the functions $\beta ^{i}=dg^{i}/d\ln M$ are the usual $\beta $%
--functions ($M$ is a mass scale) associated to the couplings $g^{i}$. To
first order in the couplings, the $\beta $--functions are given by 
\begin{equation*}
\beta ^{i}=\left( \Delta _{i}-1\right) g^{i}+o\left( g^{2}\right) ,
\end{equation*}
where $\Delta _{i}$ is the dimension of the operator $V_{i}$.

We will use the above general prescription to compute the tachyon field
effective action in the limit of large $B\rightarrow \infty $. Before doing
so, let us though discuss some issues regarding the normalization (see also 
\cite{GS}) of the path integral (\ref{eq10}), by considering the action $%
S\left( 0\right) $ at zero value of the coupling constants $g^{i}$. On one
hand, it is well known that the value of the action for a brane (maximal $25$%
--brane, since we are considering open strings with Neumann boundary
conditions) with a constant magnetic field $B_{12}=B$ is given by the
Born--Infeld value 
\begin{equation*}
S\left( 0\right) =T_{25}V_{\perp }\int d^{2}x\left( 1+B^{2}\right) ^{%
{\frac12}%
},
\end{equation*}
where $V_{\perp }$ is the volume in the perpendicular directions $y_{\perp }$
(we can think of the perpendicular directions as being compactified on a
torus) and $T_{25}$ is the tension of the $25$--brane. On the other hand, at
the point $g=0$ in parameter space clearly all beta--functions $\beta ^{i}$
vanish, and therefore $S\left( 0\right) =Z\left( 0\right) $. This implies
that the normalization for the path--integral which defines $Z\left(
g\right) $ for general values of the couplings is given by 
\begin{equation*}
\left\langle \cdots \right\rangle _{I_{0}}=\int DX\,e^{-I_{0}\left( X\right)
}\cdots \rightarrow T_{25}V_{\perp }\int d^{2}x\left( 1+B^{2}\right) ^{%
{\frac12}%
}\int D\xi \,e^{-I_{0}\left( \xi \right) }\cdots ,
\end{equation*}
where the integral over the fields $X^{a}=x^{a}+\xi ^{a}$ is given by an
integral over the zero modes $x^{a}$ and an integral over the non--zero
modes $\xi ^{a}$, normalized so that $\int D\xi e^{-I_{0}\left( \xi \right)
}\,\mathbf{1}=1$.

We now consider a generic tachyon profile given by an arbitrary function $%
t\left( x\right) $. The associated boundary operator is nothing but 
\begin{equation*}
V=t\left( X\right) .
\end{equation*}
We will show that, in the limit of large magnetic field $B\rightarrow \infty 
$, we are able to compute both the partition function $Z(t)$, and the
corresponding action $S\left( t\right) $. To this end, we first recall \cite
{SW} that, for generic functions $f_{1}\left( x\right) ,\cdots ,f_{n}\left(
x\right) $, and for generic points $\tau _{1},\cdots ,\tau _{n}$ cyclically
ordered on the boundary of the world--sheet $\Sigma $, one has that, in the
limit $B\rightarrow \infty $, 
\begin{equation*}
\left\langle f_{1}\left( X\left( \tau _{1}\right) \right) \cdots f_{n}\left(
X\left( \tau _{n}\right) \right) \right\rangle _{I_{0}}\simeq T_{25}V_{\perp
}\int d^{2}x\,B\,\left( f_{1}\star \cdots \star f_{n}\right) .
\end{equation*}
In the above equation, the $\star $--product is given as usual \cite{SW} by $%
f\star g=fg+\frac{i}{2}\theta ^{ab}\partial _{a}f\partial _{b}g+\cdots $,
with $\theta ^{12}=\theta =B^{-1}$. Therefore, in the large $B$ limit, one
can compute the partition function 
\begin{eqnarray*}
Z\left( t\right) &=&\left\langle \exp \left( -\frac{1}{2\pi }\int t\left(
X\left( \tau \right) \right) \,d\tau \right) \right\rangle _{I_{0}} \\
&\simeq &V_{\perp }T_{25}\int d^{2}x\,B\,\left( 1-t+\frac{1}{2}t\star
t+\cdots \,\right) \\
&=&V_{\perp }T_{25}\int d^{2}x\,\,B\,e^{-t}.
\end{eqnarray*}
We now rewrite the above result in terms of operators acting on a
one--particle quantum Hilbert space. This is easily done using the standard
function--operator correspondence (for example, see \cite{GMS2}), which
assigns to each function $f\left( x\right) $ an operator $Q_{f}$ such that $%
Q_{f}\cdot Q_{g}=Q_{f\star g}$ and such that $\mathrm{Tr}\left( Q_{f}\right)
=\frac{1}{2\pi }\int d^{2}x\,B\,f$. We can then rewrite the partition
function $Z$ as a function of the operator 
\begin{equation*}
T=Q_{t}
\end{equation*}
as 
\begin{equation*}
Z\left( T\right) =2\pi V_{\bot }T_{25}\mathrm{Tr}\left( e^{-T}\right) .
\end{equation*}

We now compute the action $S$ as a function of the tachyon field. This is
extremely simple in the limit of large $B$, since all the anomalous
dimensions of boundary operators go to zero \cite{SW}. To quickly realize
this fact, one must simply recall that the anomalous dimension of composite
operators comes from the logarithmic divergence of the two--point function $%
\left\langle X^{a}X^{b}\right\rangle $ on the boundary of the string
world--sheet $\Sigma $, which is proportional \cite{SW} to the so called
open string metric $G^{ab}\simeq -\left( B^{-2}\right) ^{ab}$. In the large $%
B$ limit, $G^{ab}$ vanishes, and so do all anomalous dimensions. This
implies that the $\beta $--function associated to the tachyon field is
simply given by\footnote{%
Also, no other field will be generated under RG flow, if it is not present
from the start in the boundary action. It is therefore consistent to restric
the string field space to only tachyon configurations.} 
\begin{equation*}
\beta _{t(x)}=-t(x).
\end{equation*}
Therefore, equation (\ref{eq20}) implies that the action is given by 
\begin{eqnarray*}
S\left( t\right) &=&\left( 1-\int d^{2}x\,t\left( x\right) \frac{\delta }{%
\delta t\left( x\right) }\right) Z\left( t\right) \\
&=&V_{\perp }T_{25}\int d^{2}x\,\,B\,\left( 1+t\right) e^{-t}
\end{eqnarray*}
or, in operator language, by 
\begin{equation}
S\left( T\right) =2\pi V_{\bot }T_{25}\mathrm{Tr}\left[ (1+T)e^{-T}\right] .
\label{eq30}
\end{equation}
This result could have been expected. One can in fact guess the above
equation by considering the action proposed in \cite{GSS,KMM}, and by
replacing, as usual \cite{SW}, ordinary products with $\star $--products. In
the large $B$ limit, the derivative corrections become negligible, and one
recovers (\ref{eq30}). The point of this letter is to show that
the formalism of background independent open string field theory actually
confirms this expectation.

We now wish to describe solitons of the above theory which represent a
single $23$--brane. To this end, we follow the reasoning of \cite{KMM} and
consider only tachyon configurations which are quadratic in the coordinates $%
x^{a}$ (and also, given the symmetry of the problem, rotationally invariant) 
\begin{equation*}
t(x)=c+uBr^{2},
\end{equation*}
with $r^{2}=\left( x^{1}\right) ^{2}+\left( x^{2}\right) ^{2}$. Recalling
the function--operator correspondence 
\begin{equation*}
\frac{1}{\sqrt{2\theta }}\left( x^{1}\pm ix^{2}\right) \rightarrow
a,a^{\dagger }
\end{equation*}
(where $\left[ a,a^{\dagger }\right] =1$), we quickly see that the
corresponding tachyon operator $T$ is given by 
\begin{equation*}
T=c+u\left( 2N+1\right) ,
\end{equation*}
where $N=a^{\dagger }a$ is the number operator. Therefore, the partition
function is given by 
\begin{eqnarray*}
Z &=&2\pi V_{\bot }T_{25}e^{-c-u}\mathrm{Tr}\left( e^{-2u{}N}\right) \\
&=&2\pi V_{\bot }T_{25}\frac{e^{-c-u}}{1-e^{-2u}}.
\end{eqnarray*}
Correspondingly, the action $S$ is

\begin{eqnarray*}
S &=&(1-c\frac{\partial }{\partial c}-u\frac{\partial }{\partial u})Z \\
&=&2\pi V_{\bot }T_{25}e^{-c}\frac{\left( u+1+c\right) e^{-u}+\left(
u-1-c\right) e^{-3u}}{\left( 1-e^{-2u}\right) ^{2}}.
\end{eqnarray*}
We now wish to find the stationary point of $S$ which corresponds to the $23$%
--brane solution. This will surely correspond to a saddle point, since all
branes are unstable to decay into the vacuum solution. In fact, we can first
find a \textit{maximum} of $S$ with respect to $c$ at the following value 
\begin{equation*}
c^{\star }\left( u\right) =-u\left( \frac{1+e^{-2u}}{1-e^{-2u}}\right) .
\end{equation*}
We can then follow the value of the action as a function of $c^{\star }$, $u$
\begin{equation*}
S(c^{\star }\left( u\right) ,u)=2\pi V_{\bot }T_{25}\frac{1}{1-e^{-2u}}\exp
\left( \frac{2u}{e^{2u}-1}\right) ,
\end{equation*}
which clearly decreases to the value 
\begin{equation*}
S(c^{\star }\left( u\right) ,u)\rightarrow 2\pi V_{\bot
}T_{25}=T_{23}V_{\bot }
\end{equation*}
at the saddle point obtained at $u\rightarrow \infty $, $c\rightarrow
-\infty $, with $c\simeq -u$. This shows, like previously in \cite{HKLM,KMM}%
, that the energy of the soliton is exactly that of the $23$--brane, as
expected. Moreover the saddle point is at infinite values of the couplings $%
u $ and $c$. This is again to be expected (as also noted in \cite{KMM})
since infrared fixed points of RG flows always occur at infinite value of
the couplings if one chooses coordinates in coupling space such that the $%
\beta $--functions are linear.

Some comments are in order:

\begin{enumerate}
\item  First of all, the $23$--brane soliton corresponds to the highly
singular tachyon profile $t\simeq u(-1+Br^{2})$ for $u\rightarrow \infty $.
This might, at first sight, contradict the intuition \cite{GMS2,HKLM} that
the solitons describing lower dimensional branes in the large magnetic field
limit have a characteristic size of order $\sqrt{\theta }$ -- \textit{i.e.} $%
Br^{2}\sim 1$. In the limit $u\rightarrow \infty $, since the size of the
soliton $Br^{2}\sim \frac{1}{u}$ is going to zero, one might worry that,
even in the highly non--commutative limit, derivative corrections to the
tachyon action are actually important in determining the energy of the
solitonic configuration. This is not the case, as can be seen if one
considers \cite{KMM} the (highly non--local) field redefinition $\Phi
=e^{-T/2}$. In terms of this new variable the minimum of the tachyon
potential is at $\Phi =0$ and the maximum at $\Phi =1$. Following the
reasoning of \cite{GMS2,HKLM}, we then expect the soliton to be simply given
by $\Phi =\pi _{0}$, where $\pi _{n}=\left| n\left\rangle {}\right\langle
n\right| $ is the projection onto the harmonic oscillator $n$--th state.
This is indeed the case, since a tachyon configuration $t\simeq u(-1+Br^{2})$
corresponds to an operator $T\simeq 2u{}N$, or to 
\begin{equation*}
\Phi \simeq e^{-u{}N}\rightarrow \pi _{0}
\end{equation*}
for $u\rightarrow \infty $.

\item  More generally, following the considerations of \cite{GMS2,HKLM}, and
using the intuition from the above remark, we see clearly that a solitons
with $n$ coincident $23$--branes can be described, within this framework,
with a tachyon profile which is a polynomial of degree $2n$ in the
coordinate variables. Let us illustrate this fact in the case of $n=2$. It
is simpler to proceed backwards, starting from the result of \cite{GMS2,HKLM}
\begin{equation*}
\Phi =\pi _{0}+\pi _{1}.
\end{equation*}
We can write the above equation as the limit, for $u\rightarrow \infty $, of 
\begin{equation*}
\Phi =e^{-2u{}N\left( N-1\right) }.
\end{equation*}
Therefore we conclude that the correct tachyon configuration is given by the
operator $T=4u{}N\left( N-1\right) $ or, using $r^{2}\star
r^{2}=r^{4}-\theta ^{2}$, by the function 
\begin{equation*}
t=u\left( B^{2}r^{4}-4Br^{2}+1\right) ,\,\ \ \ \ \ \ \ \ \ \ \ \ \ \ \ \ \
\left( u\rightarrow \infty \right)
\end{equation*}
which is quartic in the coordinate variables, as claimed.
\end{enumerate}

\section{Conclusions}

We have computed, using background independent open string field theory,
both the tachyon potential, as well as the explicit form and the energies of
soliton solutions corresponding to lower dimensional branes. This was
achieved by considering a closed string background with large $B_{ab}$
field. In this case, the partition function which defines the theory is
easily computed. Moreover, all anomalous dimensions vanish in the limit of
large $B_{ab}$, and the $\beta $--function flow in the space of couplings,
which is needed to compute the string action, is extremely simple, and can
be analyzed explicitly for generic tachyon profiles.

One now has a handle, within background independent open string field
theory, on the theory in two different closed string backgrounds. On the
other hand, it is conjectured \cite{S1,S2}, that the end--points of tachyon
condensations actually correspond to the same point in configuration space
of string field theory. In one description ($B_{ab}=0$) all derivative $%
\alpha ^{\prime }$ corrections count, but in the other ($B_{ab}\rightarrow
\infty $) none of them do. How this independence on the background $B_{ab}$
is achieved at the level of string field actions is still unclear, and is an
open problem for future research.

\section{Acknowledgments}

This research is supported by a European Post--doctoral Institute Fellowship.

\end{document}